\documentclass{article}

\usepackage{arxiv}

\usepackage[utf8]{inputenc} 
\usepackage[T1]{fontenc}    
\usepackage{hyperref}       
\usepackage{url}            
\usepackage{booktabs}       
\usepackage{amsfonts}       
\usepackage{nicefrac}       
\usepackage{microtype}      
\usepackage{lipsum}		
\usepackage{graphicx}
\usepackage{multirow}
\usepackage{cite}

\usepackage{amsmath}
\usepackage{chemformula}

\usepackage{epstopdf}
\usepackage{array}
\usepackage{subfigure}
\usepackage{ragged2e}
\usepackage{booktabs,makecell,multirow,tabularx}
\usepackage[ruled,linesnumbered]{algorithm2e}

\title{DeepCSNet: a deep learning method for predicting electron-impact doubly differential ionization cross sections}

\author{ Yifan Wang,~Linlin Zhong\thanks{This paper has been accepted by \href{https://doi.org/10.1088/1361-6595/ad8218}{Plasma Sources Science and Technology}} \\
	School of Electrical Engineering, Southeast University\\
	No 2 Sipailou, Nanjing, Jiangsu Province 210096, China\\
	\texttt{linlin@seu.edu.cn}\\
}

\date{July 11, 2024}

\renewcommand{\headeright}{}
\renewcommand{\undertitle}{A Preprint}

\begin{document}
\maketitle

\begin{abstract}
Electron-impact ionization cross sections of atoms and molecules are essential for plasma modelling. However, experimentally determining the absolute cross sections is not easy, and ab initio calculations become computationally prohibitive as molecular complexity increases. Existing AI-based prediction methods suffer from limited data availability and poor generalization. To address these issues, we propose DeepCSNet, a deep learning approach designed to predict electron-impact ionization cross sections using limited training data. We present two configurations of DeepCSNet: one tailored for specific molecules and another for various molecules. Both configurations can typically achieve a relative $L^2$ error less than 5\%. The present numerical results, focusing on electron-impact doubly differential ionization cross sections, demonstrate DeepCSNet's generalization ability, predicting cross sections across a wide range of energies and incident angles. Additionally, DeepCSNet shows promising results in predicting cross sections for molecules not included in the training set, even large molecules with more than 10 constituent atoms, highlighting its potential for practical applications.
\end{abstract}

\section{INTRODUCTION}
\label{sec:sec1}
\paragraph{}
Cross sections describe the scattering probability of collision process between two or more particles and are indispensable parameters in plasma modelling \cite{VA2018, cite2}. Among the collision processes driven by different particles (electron, proton, photon, etc.), electron-impact cross sections show great diversity, covering elastic collision, ionization, excitation, attachment and other cross sections, and can be complex to calculate. Current methods for calculating electron-impact cross sections fall into two categories: solving the inverse swarm problem based on experimentally measured electron swarm parameters \cite{cite3, cite4} and directly calculating cross sections based on ab initio methods \cite{cite5, cite6, cite7, cite8, cite9, cite10, cite11, cite12, cite13, cite14, cite15}. Christophorou used electron swarm parameters to calculate and analyze the electron-impact cross sections of fluorocarbon molecules, covering ionization, decomposition, attachment, and excitation cross sections \cite{cite3}. Similar approaches were applied by Franck et al to acquire electron-impact cross sections of new refrigerant gas \cite{cite4}. However, results obtained by this method show insufficient consistency due to different experimental conditions, which imposes significant limitations. As for ab initio methods, numerous theoretical methods have been developed to obtain cross sections by accurately solving the Schrödinger equation, such as the close-coupling method with its variants \cite{cite5, cite6, cite7, cite8} by diagonalizing the target Hamiltonian, and R-matrix method \cite{cite9, cite10} by dividing the space into an inner region and an outer region. However, these theoretical methods generally exhibit great computational complexity. Kim and Rudd \cite{cite11}, as well as Deutsch and Märk \cite{cite12}, respectively proposed the Binary-Encounter-Bethe (BEB) method and the Deutsch-Märk (DM) method to simplify the calculations. Many works have obtained ionization cross sections for various molecules using the BEB and DM models \cite{cite13, cite14, cite15}. Nevertheless, the computational cost still remains high, especially for large molecules. These methods only provide total cross sections rather than differential cross sections (DCS), which are necessary parameters for many plasma models \cite{cite16}. Limited by the measurement resolutions, uncertainties and calculation methods, there is a lack of comprehensive DCS data across various energies and incident angles of certain molecules \cite{cite17}, especially for ionization DCS. In the process of electron-impact ionization, two electrons, called the scattered electron and the ejected electron, are involved. By taking the ejected energy into consideration, doubly differential cross section (DDCS) data can be measured or calculated \cite{cite18, cite19}, helping investigate collision dynamics and particle systems \cite{cite20}, and DCS can be directly obtained by integrating DDCS.

\paragraph{}
Recently, with the development of artificial intelligence (AI), researchers have begun integrating AI with electron-impact cross section calculations. Zhong et al. \cite{cite21} proposed a support vector machine (SVM)-based method to predict electron-impact cross sections for large molecules using data from smaller molecules for training, although the calculation of smaller molecules still consumed significant time. Jetly et al. \cite{cite22} and Stokes et al. \cite{cite23, cite24} both treated the calculation of cross sections as an inverse problem of the Boltzmann equation. The former predicted cross section data from the solutions of the Boltzmann equation through neural networks. The latter estimated elastic momentum transfer and ionization cross sections from electron swarm parameters using deep neural networks. However, this method does not consider intrinsic information about the particles such as molecular compositions and the uniqueness of solution is not always guaranteed. A more direct method, proposed by Harris et al. \cite{cite25, cite26}, predicted proton-impact and electron-impact cross sections based on molecular formulas using neural networks. However, this method could only output cross section data at fixed incident angles and exhibited poor generalization, causing a typical difference from test data by 10\% to 30\%.

\paragraph{}
This work proposes a deep learning-based method called Deep Cross Section Network (DeepCSNet) for predicting electron-impact ionization cross sections, specifically doubly differential ionization cross sections and total ionization cross sections. DeepCSNet includes two specific configurations, the single molecule configuration, which predicts cross sections for a specific molecule, and the multi-molecule configuration, which predicts cross sections for molecules within a single model. For DeepCSNet in both configurations, the performance and the generalization abitlity across energies, incident angles and molecules are investigated. Section 2 details the framework of DeepCSNet. Section 3 presents the numerical results for the single molecule configuration, while Section 4 covers the results for the multi-molecule configuration. Finally, this work is concluded in Section 5.

\section{METHOD}
\label{sec:sec2}
\subsection{DeepONet}
\label{sec:sec2.1}
\begin{figure}
	\centering
	\includegraphics[width=9cm]{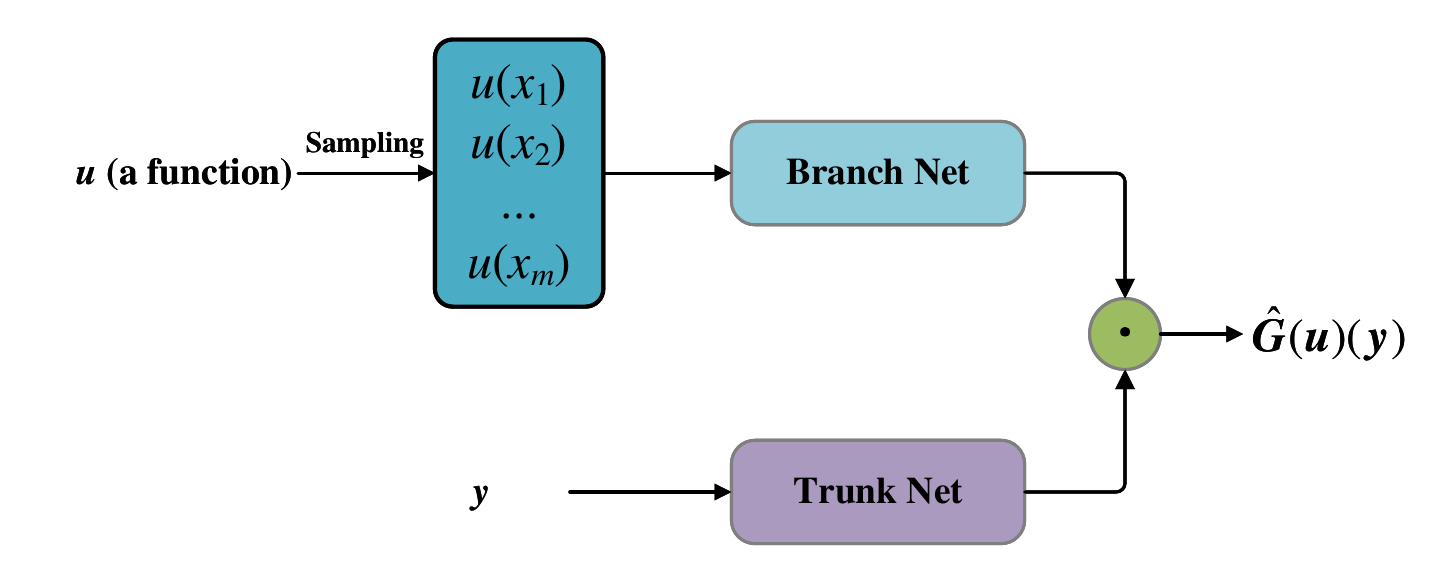}
	\caption{The framework of vanilla DeepONet}
	\label{fig:fig1}
\end{figure}

\paragraph{}
The Deep Operator Network (DeepONet) is a neural network framework first proposed by Lu et al. \cite{cite27} to learn various operators. DeepONet consists of two subnets: a branch net to encode the input function u and a trunk net to encode the temporal-spatial domain y of the output. The output of DeepONet $\hat{G}(u)(y)$ is the dot product of the outputs of the branch net and the trunk net, as shown in Figure \ref{fig:fig1}.

\paragraph{}
The branch net and the trunk net are both dense neural networks (DNNs), so the output can be described as:

\begin{equation}
\label{equ:equ1}
	\hat{G}(u)(y) = \text{NN}_{\text{branch}}(u) \cdot \text{NN}_{\text{trunk}}(y)
\end{equation}

\begin{equation}
\label{equ:equ2}
	\text{NN}(z^0) = z^k = \sigma(\mathbf{w}_k z^{k-1} + \mathbf{b}_k), \quad k = 1, 2, \dots, l
\end{equation}

where $\hat{G}$ is the approximation to nonlinear operator $G$, $\text{NN}(\cdot)$ presents the calculation by the corresponding neural network, $z^{0}$ is the input of DNN, $l$ is the layer number of DNN, $\sigma$ is the activation function, $\mathbf{w}_k$ and $\mathbf{b}_k$ are the weights and bias of the $k$-th layer.

\paragraph{}
Supported by the Universal Approximation Theorem \cite{cite28}, DeepONet can theoretically approximate any nonlinear operator when it converges according to the following loss function:

\begin{equation}
\label{equ:equ3}
	L=\frac{1}{N-M} \sum_{i=1}^{N} \sum_{j=1}^{M}\left(\hat{G}\left(u_{i}\right)\left(y_{j}\right)-G\left(u_{i}\right)\left(y_{j}\right)\right)^{2}
\end{equation}

where $L$ is the loss function, $N$ and $M$ are the number of sampling points for function $u$ and coordinates $y$, respectively, and $\hat{G}(u)(y)$ and $G(u)(y)$ are the predicted results and label values, respectively. Once trained, DeepONet can directly predict $G$ for different input functions $u$ and coordinates $y$, demonstrating strong generalization performance.

\subsection{DeepCSNet}
\label{sec:sec2.2}
\begin{figure}
	\centering
	\includegraphics[width=9cm]{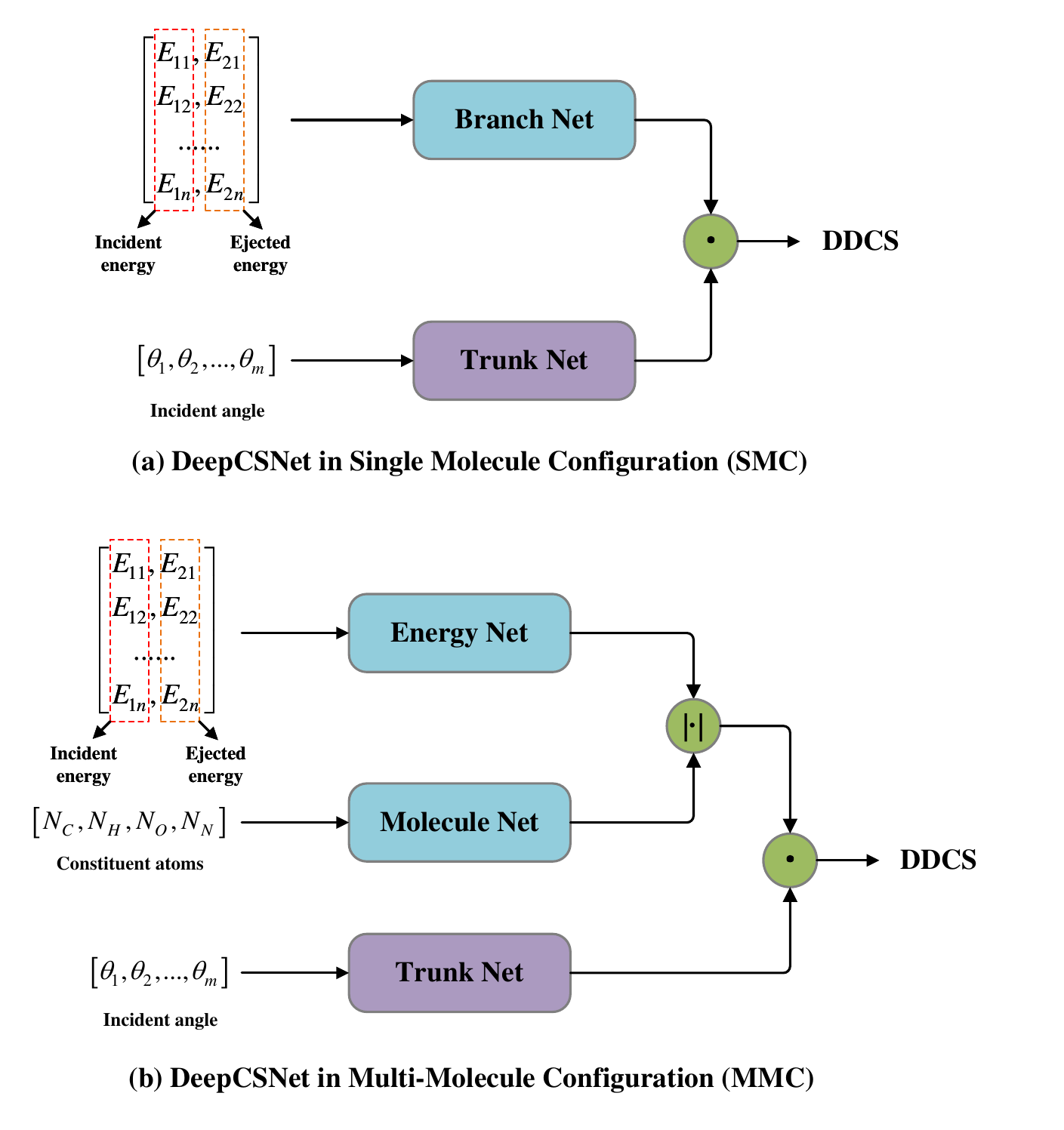}
	\caption{The framework of DeepCSNet for DDCS. (a) SMC without molecule information informed. (b) MMC with constitute atoms informed}
	\label{fig:fig2}
\end{figure}

\paragraph{}
This work considers DDCS in the process of electron-impact ionization. Therefore, the calculation of cross sections can be treated as a nonlinear operator, where the inputs are incident angles, incident energies, and ejected energies, and the output is the DDCS. Based on this consideration, the framework of Deep Cross Section Network (DeepCSNet) can be derived as shown in Figure \ref{fig:fig2}. Two specific configurations are proposed: a single molecule configuration (SMC) and a multi-molecule configuration (MMC).

\paragraph{}
2.2.1 Single molecule configuration

\paragraph{}
In this configuration, the model predicts DDCS for one specific molecule. Therefore, the model only takes incident and ejected energies and incident angles as input, and no molecular information is required for the model. The DDCS is considered as a function distribution on the axis of incident angles, while the incident energies and the ejected energies are treated as parameters of this function. Consequently, the branch net takes incident energies and ejected energies as input and the trunk net takes incident angles as input.

\paragraph{}
The predicted DDCS can be written as:

\begin{equation}
\label{equ:equ4}
	\hat{Q}(E_i, E_j, \theta) = \text{NN}_{\text{branch}}(E_i, E_j) \cdot \text{NN}_{\text{trunk}}(\theta)
\end{equation}

Where $\hat{Q}$  is the predicted DDCS, NN$(\cdot)$ presents the calculation by the corresponding neural network, $E_1$ and $E_2$ are the incident and ejected energies, respectively, and $\theta$ is the incident angle.

\paragraph{}
2.2.2 Multi-molecule configuration

\paragraph{}
In this configuration, one trained model can predict the DDCS for different molecules. The trunk net remains the same as SMC. However, the branch net is separated into two subnets for better performance due to the significant difference in magnitude between the number of atoms and the energies. The energy net takes the incident and ejected energies as input, while the molecule net takes the number of consituent atoms as input. Before dot multiplication with the output of the trunk net, the outputs of the energy net and the molecule net are concatenated. For molecules composed of C, H, O and N, the predicted DDCS can be written as:

\begin{equation}
\label{equ:equ5}
	\hat{Q}(E_i, E_j, \theta, N_C, N_H, N_O, N_N) = \left| \text{NN}_{\text{energy}}(E_i, E_j) \cdot \text{NN}_{\text{molecule}}(N_C, N_H, N_O, N_N) \right| \cdot \text{NN}_{\text{trunk}}(\theta)
\end{equation}
Where |$\cdot$| denotes the concatenation operation, and $N_C$, $N_H$, $N_O$, and $N_N$ are the numbers of carbon, hydrogen, oxygen, and nitrogen atoms, respectively.

\section{SINGLE MOLECULE CONFIGURATION RESULTS}
\label{sec:sec3}
\subsection{Model training and testing}
\label{sec:sec3.1}
\paragraph{}
DDCS for 10 different molecules (O$_2$, N$_2$, H$_2$, CH$_4$, NH$_3$, H$_2$O, CO, C$_2$H$_2$, NO, CO$_2$) composed of C, H, O and N are used to train and test the model separately, along with 5 noble gases (He, Ne, Ar, Kr, Xe). All the data are from the experimental measurements of Opal et al \cite{cite29}. For He, O$_2$ and N$_2$, 495 sets of data at 9 incident angles from 30° to 150° are included for each molecule. Among them, 395 sets are used as training set and 100 sets as test set. The incident energy ranges from 50 to 2000 eV, and the ejected energy ranges from 4.13 to 205 eV. For the other molecules, 85 sets of data at 9 incident angles are included for each molecule. Among them, 60 sets are used as training set and 25 sets as test set. The incident energy is set to 500 eV \cite{cite29}, and the ejected energy ranges from 4.13 to 205 eV. Before training, the DDCS and the incident energies are converted to logarithmic values due to their large span, and then normalized to [0.05, 0.95]. The ejected energies are normalized to [0.05, 0.95] as well \cite{cite26}.

\begin{figure}
	\centering
	\includegraphics[width=9cm]{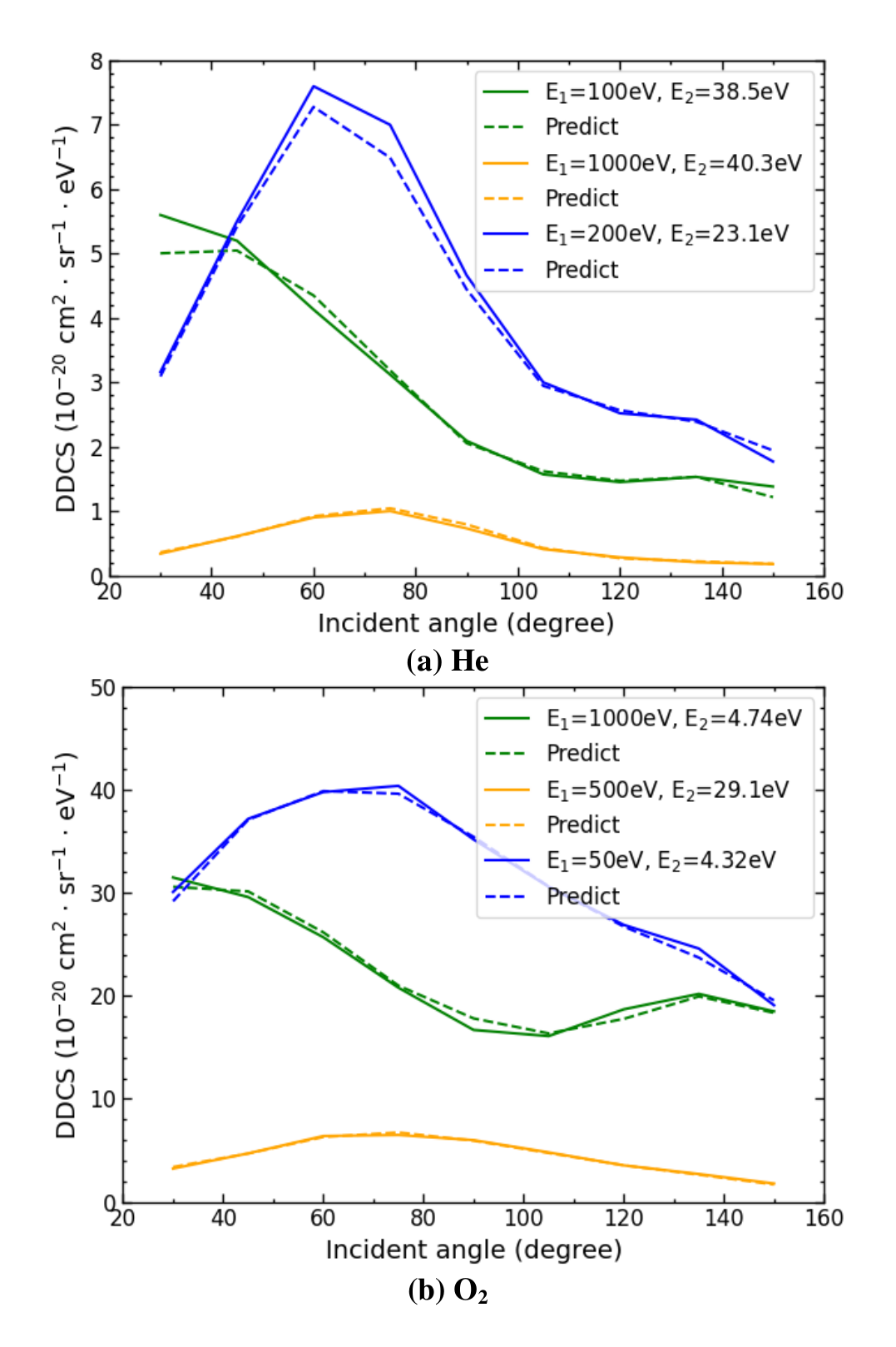}
	\caption{SMC predictions of DDCS for He and O$_2$}
	\label{fig:fig3}
\end{figure}

\paragraph{}
The branch net and the trunk net both consist of 3 hidden layers with 80 neurons per layer. The sigmoid activation function \cite{cite30} is used for both networks, and the Adam algorithm \cite{cite31} is applied for optimization. After training, the model is tested on the test set. Parts of the results are given in Figure \ref{fig:fig3}, and Table \ref{tab:tab1} presents the mean relative $L^2$ error for each molecule.

\begin{table}[htbp]
	\centering
	\caption{Mean relative $L^2$ error of predictions to DDCS in SMC}
	\label{tab:tab1}
	\begin{tabular}{lllll}
	\cline{1-4}
		Molecule & Mean relative $L^2$ error & Molecule & Mean relative $L^2$ error &  \\ \cline{1-4}
		He       & 0.045                  & CH$_4$      & 0.047                  &  \\
		Ne       & 0.024                  & NH$_3$      & 0.055                  &  \\
		Ar       & 0.058                  & H$_2$O      & 0.039                  &  \\
		Kr       & 0.12                   & CO       & 0.040                  &  \\
		Xe       & 0.076                  & C$_2$H$_2$     & 0.035                  &  \\
		O$_2$       & 0.035                  & NO       & 0.044                  &  \\
		N$_2$       & 0.040                  & CO$_2$      & 0.025                  &  \\
		H$_2$       & 0.023                  &          &                        &  \\ \cline{1-4}
	\end{tabular}
\end{table}

\paragraph{}
The results show that DeepCSNet in SMC can accurately predict DDCS for a set of molecules, even with limited training data. The predictions for Kr exhibit a slight increase in error, likely due to the relatively larger span of DDCS values for Kr. The visualizations demonstrate that the different patterns in various incident and ejected energies can be well reconstructed.

\subsection{Generalization to unknown energies}
\label{sec:sec3.2}
\paragraph{}
The generalization performance for unknown energies is tested for the single molecule configuration. In this work, 'unknown' refers to 'not included in training set'. DDCS for He and O$_2$ are specifically considered, as the data of He and O$_2$ contains different incident energies.

\begin{figure}
	\centering
	\includegraphics[width=9cm]{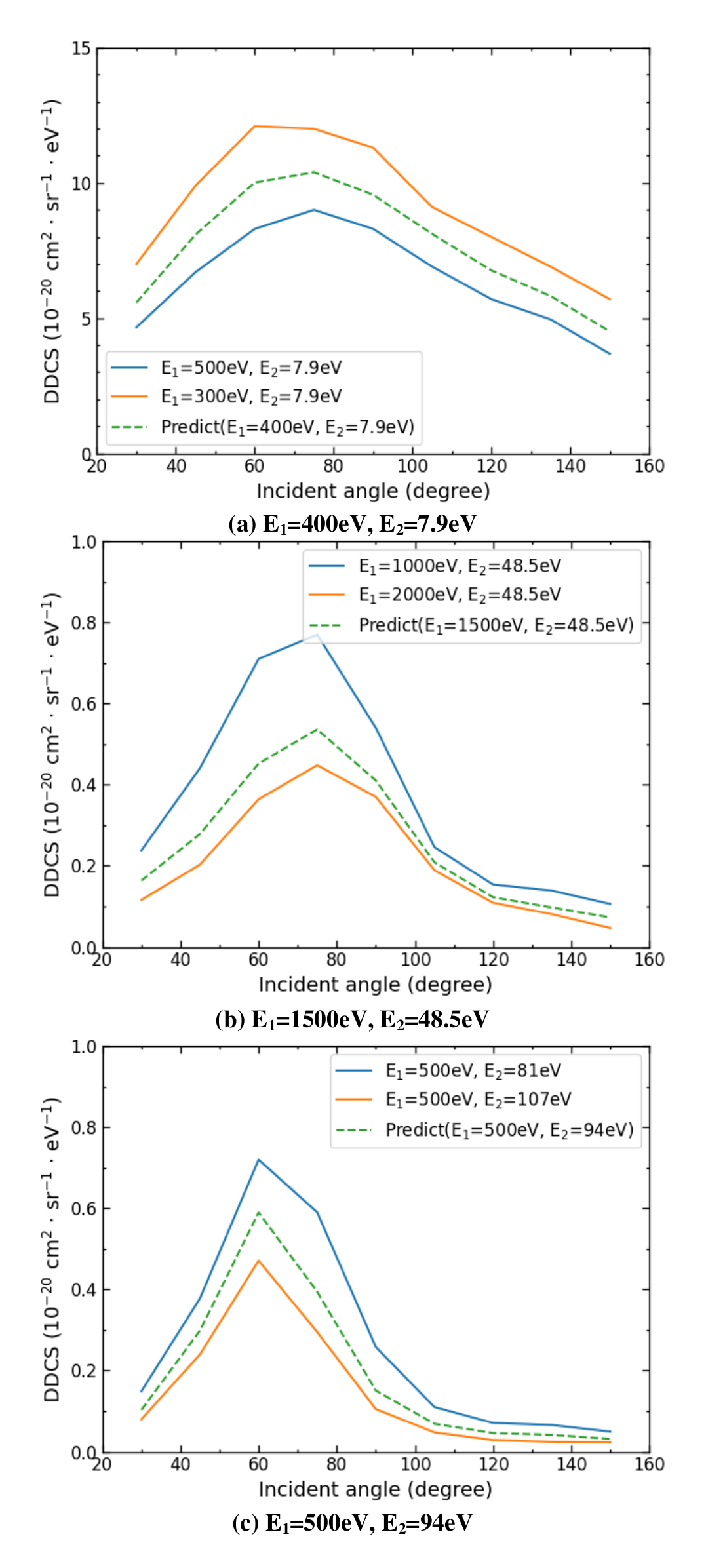}
	\caption{SMC predictions of DDCS for He at unknown energies within training range}
	\label{fig:fig4}
\end{figure}

\begin{figure}
	\centering
	\includegraphics[width=9cm]{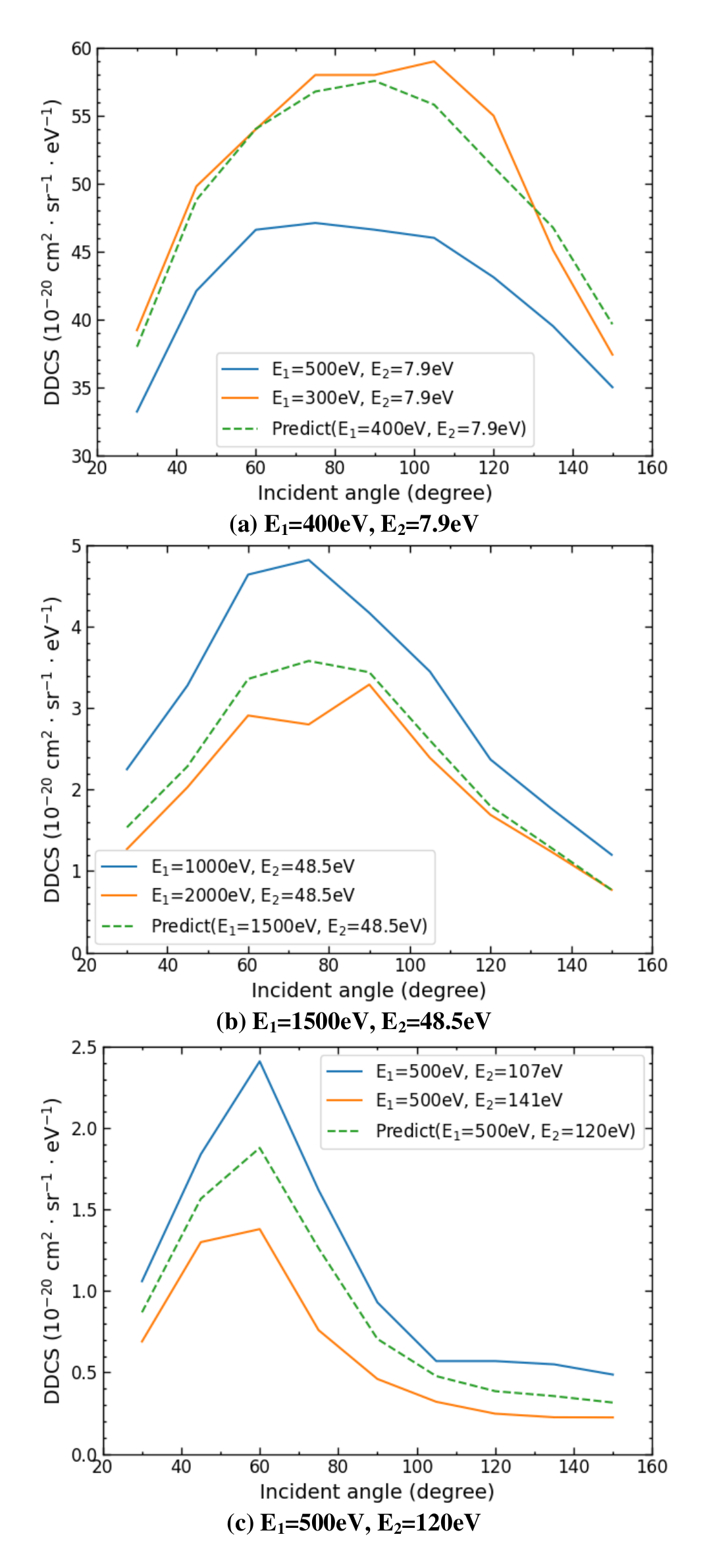}
	\caption{SMC predictions of DDCS for O$_2$ at unknown energies within training range}
	\label{fig:fig5}
\end{figure}

\paragraph{}
The generalization to unknown energies within the training range (incident energy from 50 to 2000eV and ejected energy from 4.13 to 205 eV) is first tested, and parts of the results are given in Figure \ref{fig:fig4} and \ref{fig:fig5}. The results indicate that DeepCSNet can well predict the trends of DDCS at unknown incident and ejected energies. Notably, the predictions for O$_2$ show that DeepCSNet can learn and distinguish the changes in DDCS at higher or lower energies, supplementing the change process in Figure \ref{fig:fig5}(b) and Figure \ref{fig:fig5}(c).

\begin{figure}
	\centering
	\includegraphics[width=9cm]{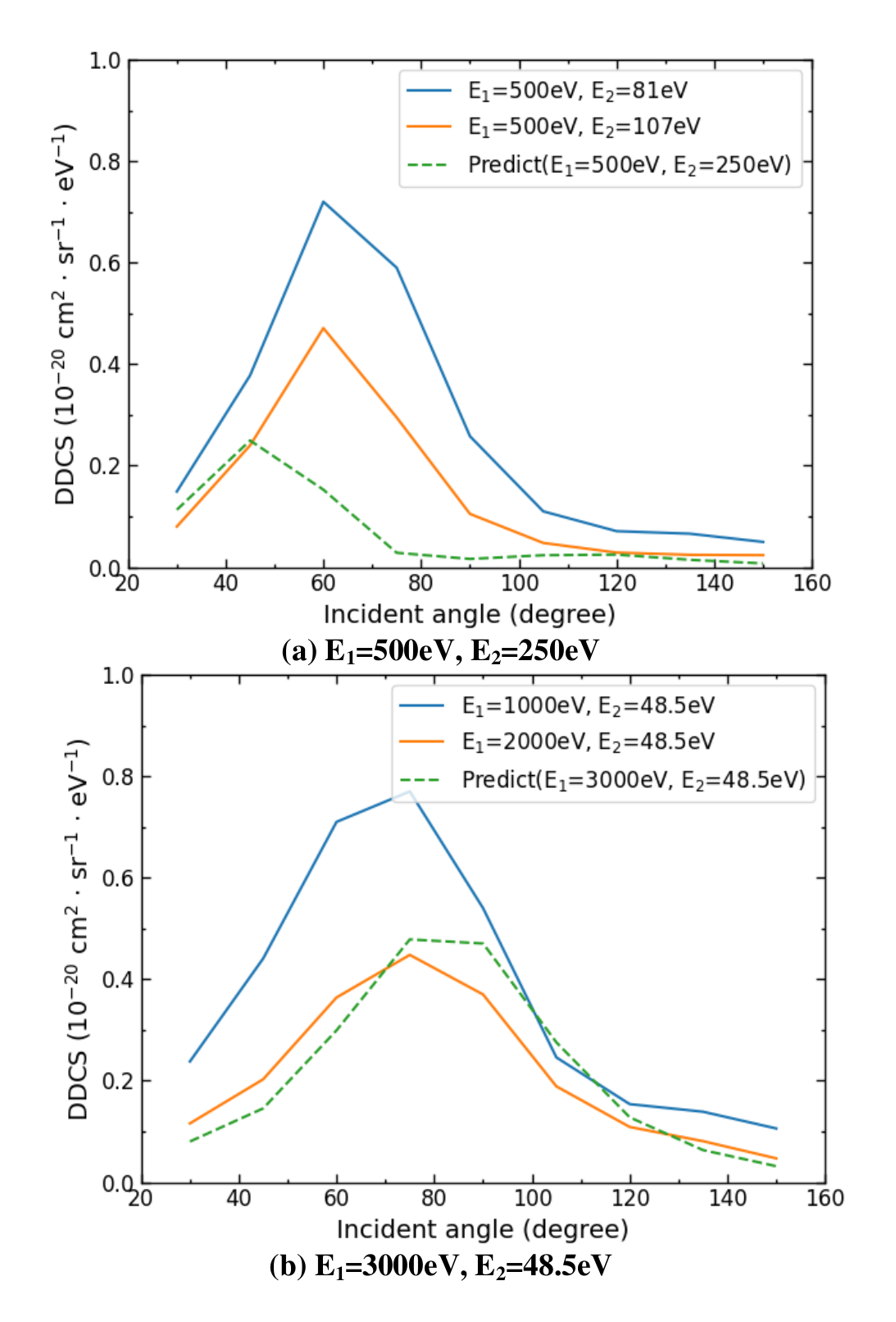}
	\caption{SMC predictions of DDCS for He at unknown energies outside training range}
	\label{fig:fig6}
\end{figure}

\begin{figure}
	\centering
	\includegraphics[width=9cm]{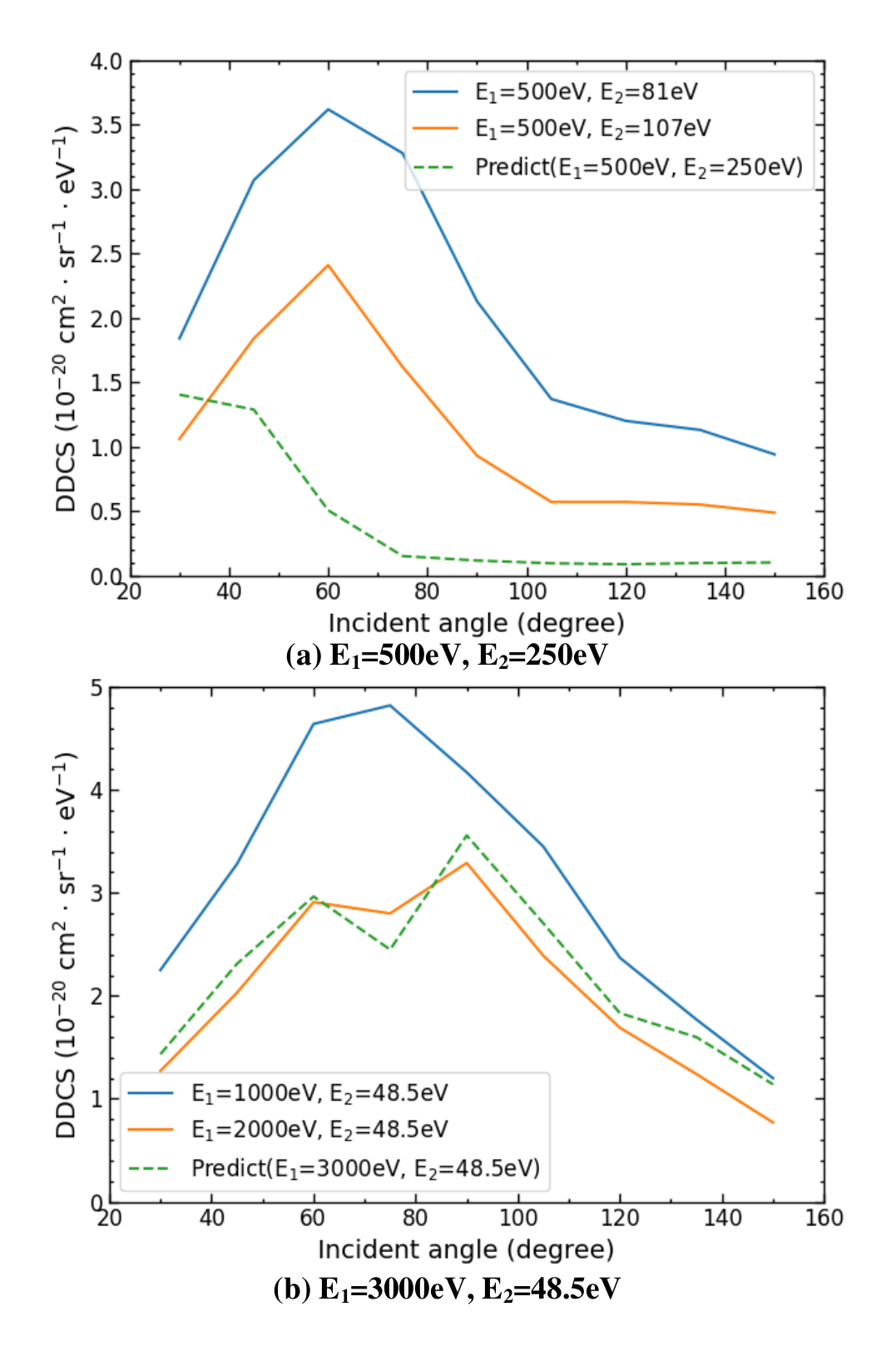}
	\caption{SMC predictions of DDCS for O$_2$ at unknown energies outside training range}
	\label{fig:fig7}
\end{figure}

\paragraph{}
The generalization to unknown energies outside the training range is analyzed as well. Figure \ref{fig:fig6} and \ref{fig:fig7} display parts of the results and prove that DeepCSNet can predict the trends of DDCS when dealing with incident and ejected energies outside the training range. However, at certain incident angle points, DeepCSNet may show a slight decrease in accuracy, especially when DDCS changes minimally across different energies. Still, the changes in DDCS can be predicted even when the energies are outside the training range, which is shown in Figure \ref{fig:fig7}(b).

\subsection{Generalization to unknown angles}
\label{sec:sec3.3}
\paragraph{}
Next, the generalization performance to unknown angles is tested. DDCS for He and O$_2$ are specifically considered.

\begin{figure}
	\centering
	\includegraphics[width=9cm]{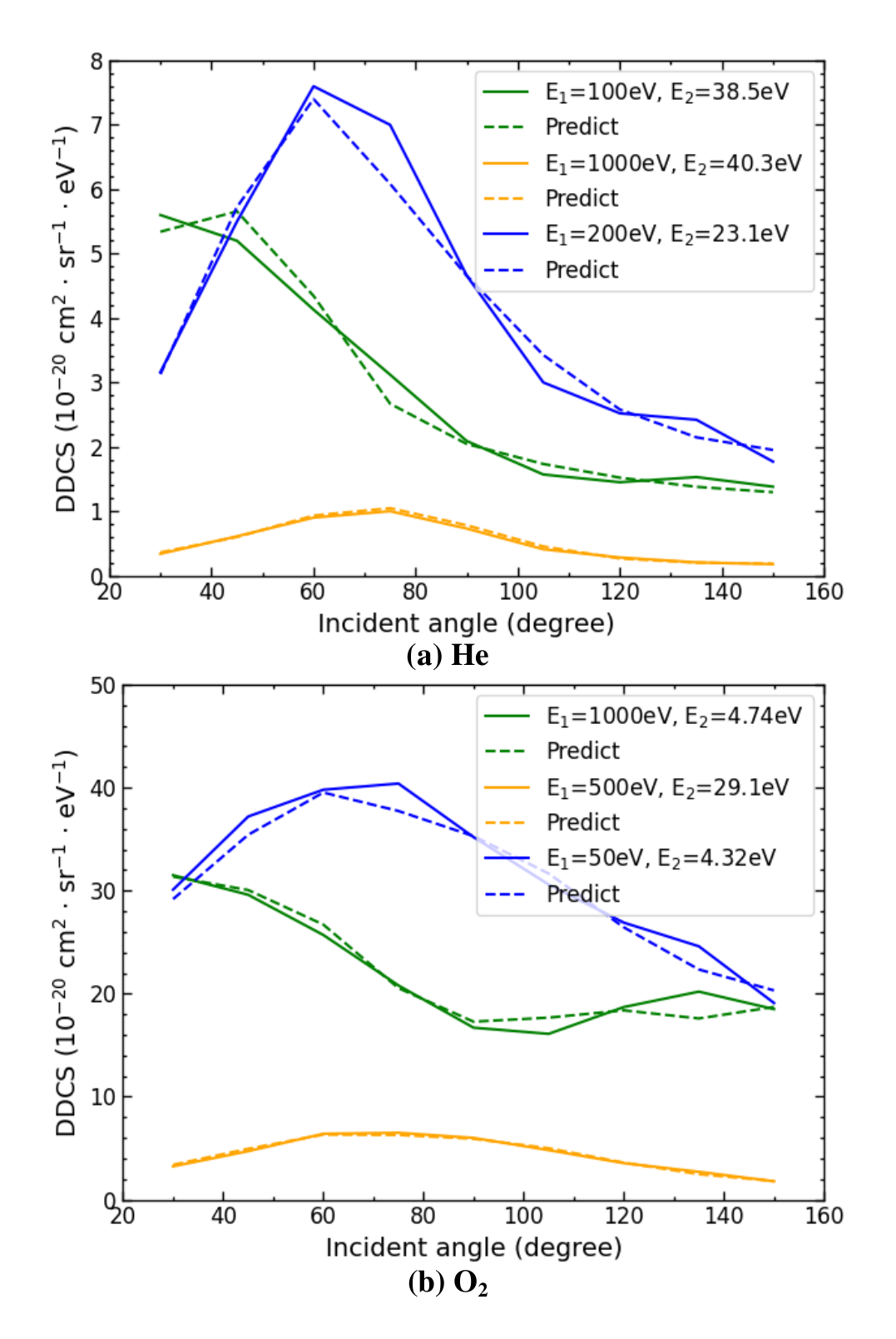}
	\caption{SMC predictions of DDCS at unknown angles within training range}
	\label{fig:fig8}
\end{figure}

\paragraph{}
The model is trained with data \cite{cite29} at 5 incident angles from 30° to 150°. The test is conducted with data at 9 incident angles within the same range, and the results are shown in Figure \ref{fig:fig8}. These results indicate that DeepCSNet can predict DDCS at unknown angles with reasonable accuracy.

\begin{figure}
	\centering
	\includegraphics[width=9cm]{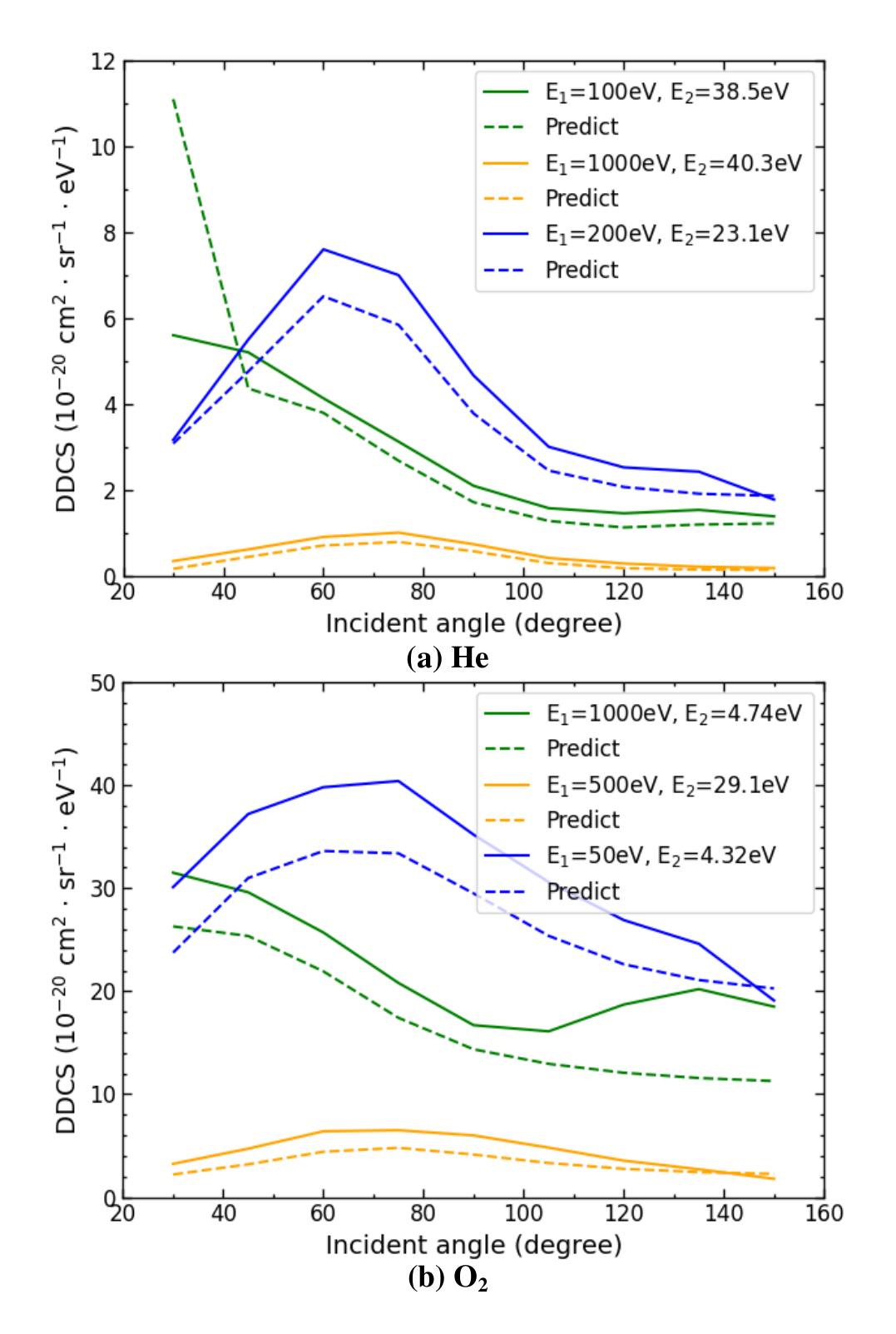}
	\caption{SMC predictions of DDCS at unknown angles outside training range}
	\label{fig:fig9}
\end{figure}

\paragraph{}
Subsequently, the model is trained with data at 7 incident angles from 45° to 135°. The test is then conducted with data at 9 incident angles from 30° to 150° to investigate the generalization performance at incident angles outside the training range. The results are illustrated in Figure \ref{fig:fig9}. In this test, DeepCSNet can still predict the overall trends of DDCS, but the predictions exhibit deviations from the data.

\section{MULTI-MOLECULE CONFIGURATION RESULTS}
\label{sec:sec4}
\paragraph{}
This section provides a comprehensive analysis of DeepCSNet in MMC. Emphasis is placed on test performance and generalization ability, particularly the model’s ability to generalize to unknown molecules.

\subsection{Model training and testing}
\label{sec:sec4.1}
\paragraph{}
DDCS for 10 different molecules (O$_2$, N$_2$, H$_2$, CH$_4$, NH$_3$, H$_2$O, CO, C$_2$H$_2$, NO, CO$_2$) composed of C, H, O, and N are used to train and test the model. The training set contains 1400 sets of data at 9 incident angles, and the test set contains 266 sets of data at 9 incident angles. The same data preprocessing strategy and training setup as before are used. The energy net and the molecule net both consists of 3 hidden layers with 40 neurons per layer, while the trunk net remains the same.

\paragraph{}
The effect of separating the branch net is investigated. A multi-molecule model with a single branch net and one with two separated branch nets are trained and tested, repectively. The mean relative $L^2$ error of the former is 0.057, and that of the latter is 0.049. Additionally, the separation of the branch net slightly reduces the scale of network parameters. Therefore, constructing DeepCSNet with two separate branch nets is more reasonable.

\begin{figure}
	\centering
	\includegraphics[width=9cm]{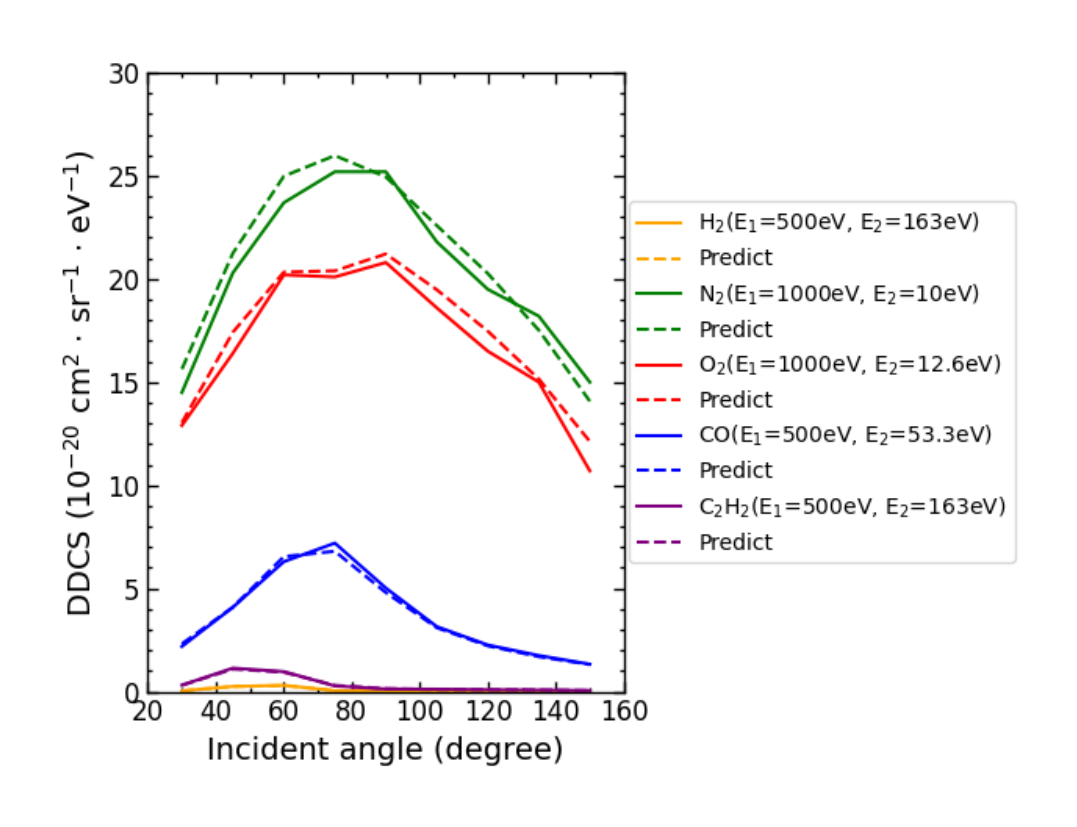}
	\caption{MMC predictions of DDCS for different molecules}
	\label{fig:fig10}
\end{figure}

\paragraph{}
Figure \ref{fig:fig10} shows parts of the predictions of DeepCSNet. The visualizations demonstrate that the varying patterns such as curve shapes, the position of the peak and the feature of two peaks can be well predicted. The results indicate that DeepCSNet in both configurations can predict DDCS for different molecules with or without the information of molecular composition, even with limited training data as few as 60. DeepCSNet in MMC effectively handles the variation in data quantity for different molecules, demonstrating potential for practical applications.

\subsection{Generalization to unknown energies}
\label{sec:sec4.2}
\paragraph{}
The generalization performance to unknown energies is tested for the multi-molecule configuration.

\begin{figure}
	\centering
	\includegraphics[width=9cm]{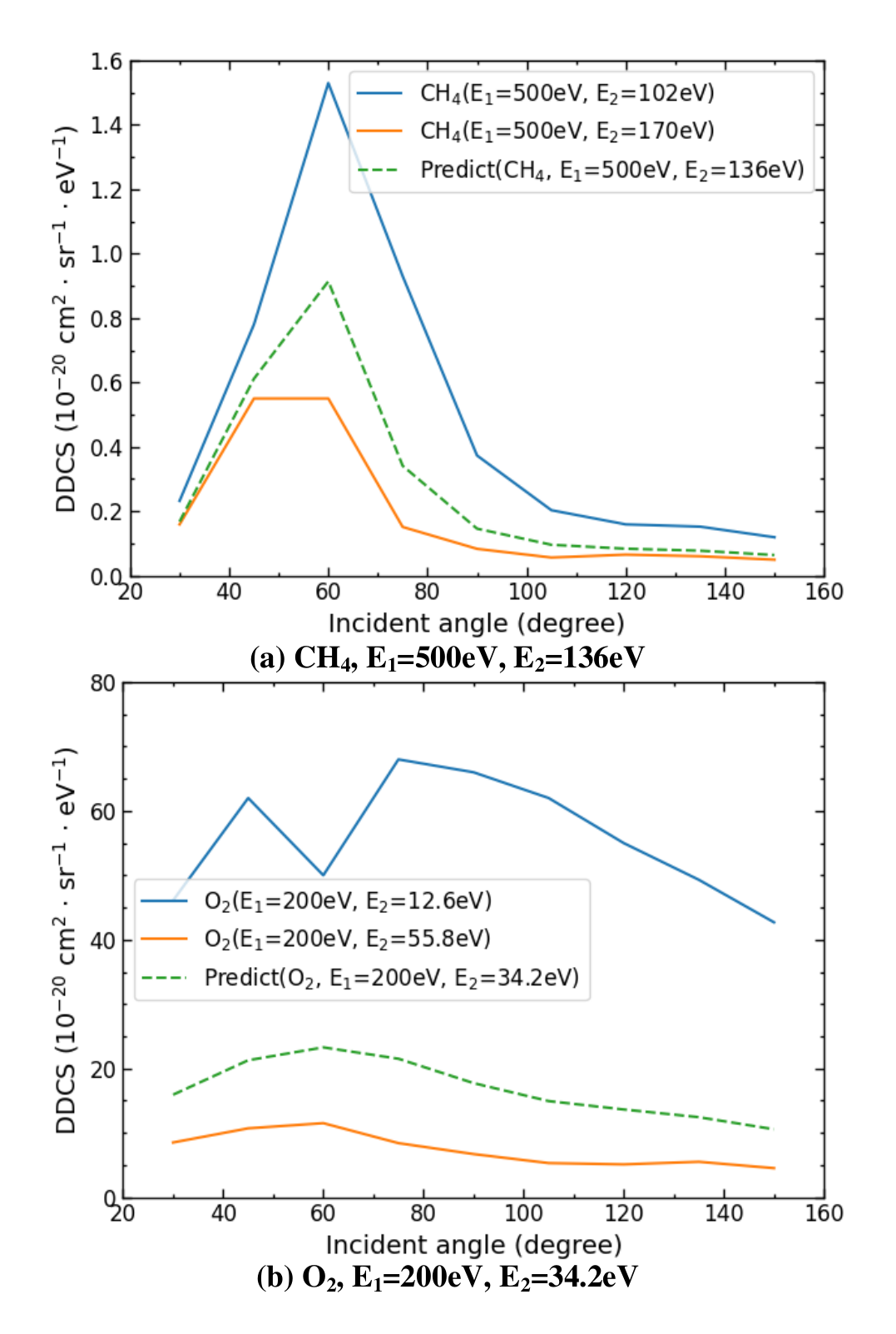}
	\caption{MMC predictions of DDCS at unknown energies within training range}
	\label{fig:fig11}
\end{figure}

\paragraph{}
First, the generalization to unknown energies within the training ranges (incident energy from 50 to 2000eV and ejected energy from 4.13 to 205 eV) is tested, and selected results are given in Figure \ref{fig:fig11}. The results show that DeepCSNet can predict the trends of DDCS for different molecules at unknown incident and ejected energies. The noticeable plateau and drop in DDCS in Figure \ref{fig:fig11}(a) and \ref{fig:fig11}(b) indicate that DeepCSNet can distinguish between energies or identify outlier data in the dataset, depending on the cause of the plateau and drop.

\begin{figure}
	\centering
	\includegraphics[width=9cm]{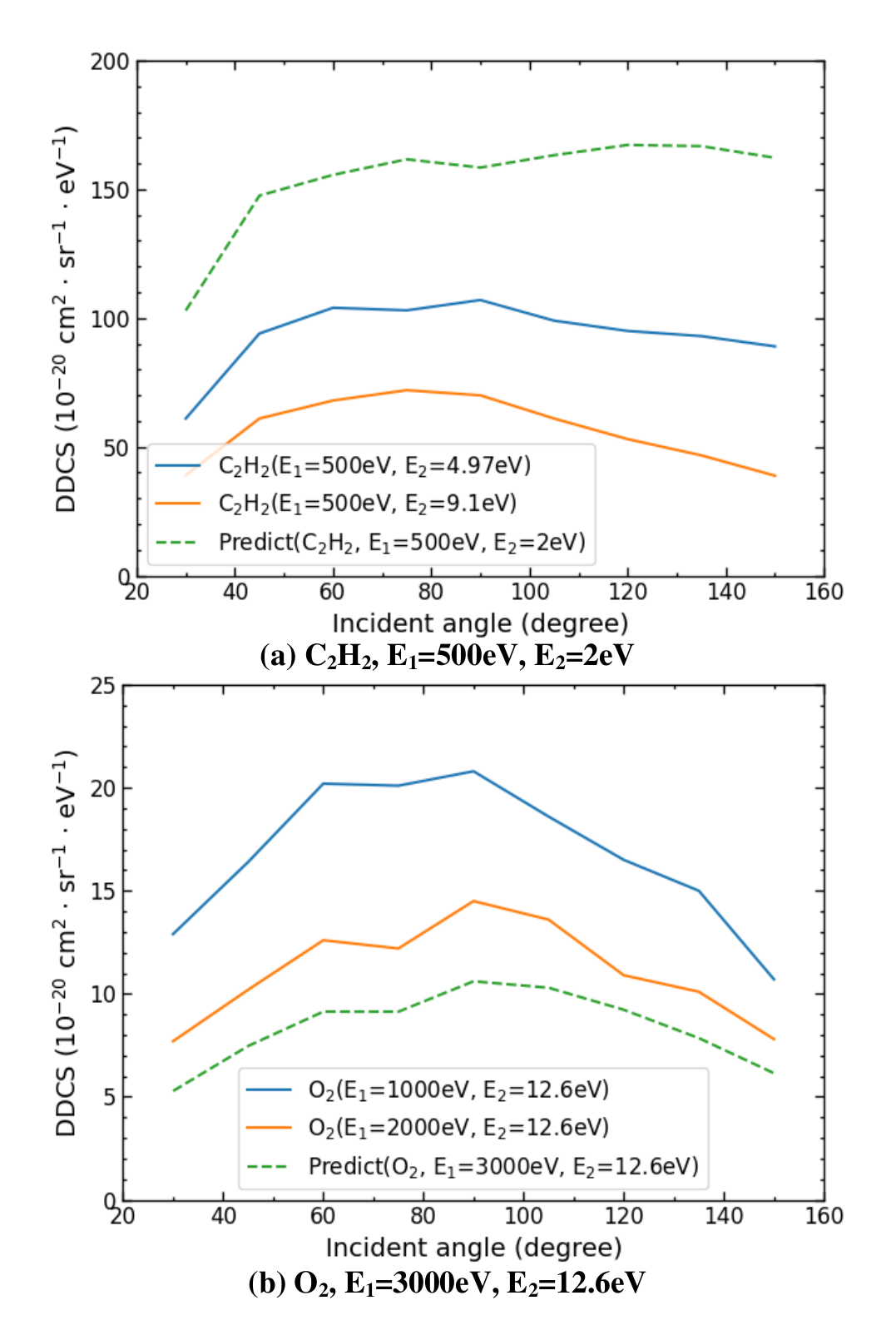}
	\caption{MMC predictions of DDCS at unknown energies outside training range}
	\label{fig:fig12}
\end{figure}

\paragraph{}
Then, the generalization to unknown energies outside the training ranges are tested as well. Figure \ref{fig:fig12} gives parts of the results. Again, DeepCSNet can predict the trends of DDCS for different molecules when dealing with incident and ejected energies outside the training range. The results also demonstrate that DeepCSNet in MMC performs better than in SMC when the incident and ejected energies are outside the training range. It accurately captures the rapid drop in DDCS at small incident angles in Figure \ref{fig:fig12}(a) and the slight twist in DDCS around 80° in Figure \ref{fig:fig12}(b). This improvement in performance may stem from the richer information in the training set, which contains different molecules rather than one single molecule.

\subsection{Generalization to unknown angles}
\label{sec:sec4.3}
\begin{figure}
	\centering
	\includegraphics[width=9cm]{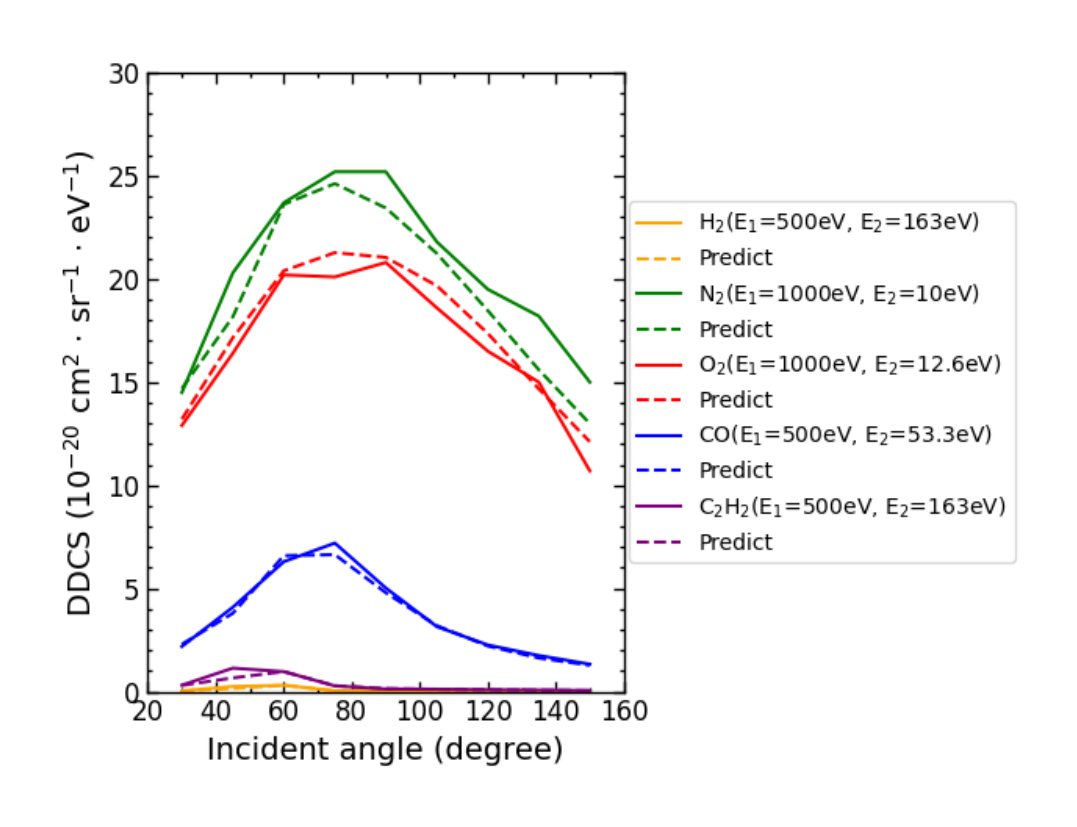}
	\caption{MMC predictions of DDCS for different molecules at unknown angles within training range}
	\label{fig:fig13}
\end{figure}

\paragraph{}
The generalization performance to unknown angles is subsequently tested. The model is trained with data at 5 incident angles from 30° to 150° \cite{cite29}. The test is conducted with data at 9 incident angles, and parts of the results are shown in Figure \ref{fig:fig13}. The mean relative $L^2$ error on test set is 0.091. These results demonstrate that DeepCSNet can predict DDCS at unknown angles with small deviations in MMC.

\begin{figure}
	\centering
	\includegraphics[width=9cm]{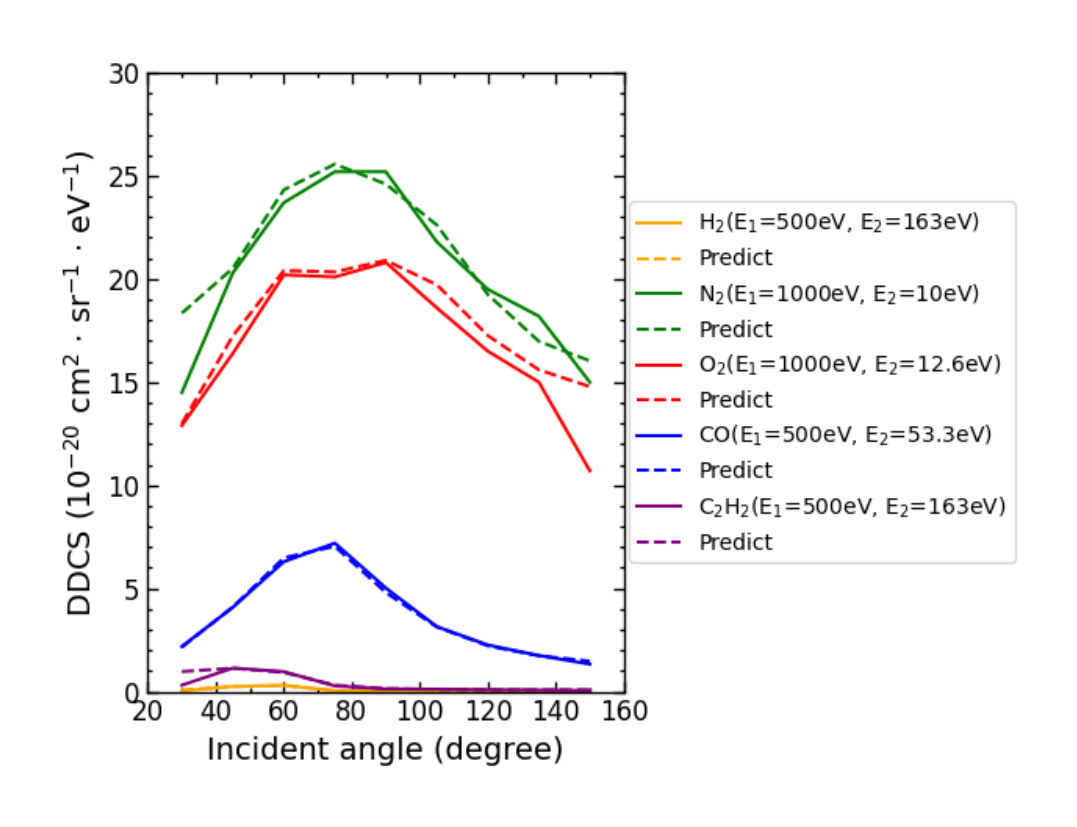}
	\caption{MMC predictions of DDCS for different molecules at unknown angles outside training range}
	\label{fig:fig14}
\end{figure}

\paragraph{}
Next, the model is trained with data at 7 incident angles from 45° to 135°. The test is conducted with data at 9 incident angles from 30° to 150°, and the results are shown in Figure \ref{fig:fig14}. The mean relative $L^2$ error on test set is 0.085. These results indicate that, in most cases, DeepCSNet can accurately predict DDCS at unknown incident angles outside the training range. Even when predictions show relatively large deviations, the trends remain correct. Again, the MMC performs better than the SMC when dealing with unknown incident angles outside the training range, demonstrating the generalization ability of DeepCSNet.

\subsection{Generalization to unknown molecules}
\label{sec:sec4.4}
\paragraph{}
Finally, the generalization ability to unknown molecules is tested, which is quite important for the multi-molecule configuration. The model is trained on the data for O$_2$, H$_2$, N$_2$ and CO$_2$. The model is then refined and tested on the data for CO. The test set for CO consists of 25 sets of data, and the refine set increases from 0 to 60, where 0 means directly predicting DDCS for CO without refinement. Both models, one with a single branch net and one with two separate branch nets, are trained and tested. The mean relative $L^2$ error are listed in Table \ref{tab:tab2}.

\paragraph{}
The results show that although a model with a single branch net can achieve lower error when more data are used for refinement, it cannot directly generalize to unknown molecules. In contrast, DeepCSNet can effectively generalize to unknown molecules after refinement with only 10 sets of data. Even when used directly for prediction, DeepCSNet maintains acceptable accuracy.

\begin{table}
	\centering
	\caption{Mean relative $L^2$ error for unknown molecules}
	\label{tab:tab2}
	\begin{tabular}{lllll}
	\cline{1-3}
	Scale of refine set & Mean relative $L^2$ error of single branch   net & Mean relative $L^2$ error of separate branch   nets (DeepCSNet) &  &  \\ \cline{1-3}
	60                  & 0.043                                         & 0.046                                                        &  &  \\
	40                  & 0.051                                         & 0.053                                                        &  &  \\
	20                  & 0.067                                         & 0.064                                                        &  &  \\
	10                  & 0.13                                          & 0.068                                                        &  &  \\
	0                   & 0.38                                          & 0.10                                                         &  &  \\ \cline{1-3}
	\end{tabular}
\end{table}

\begin{figure}
	\centering
	\includegraphics[width=9cm]{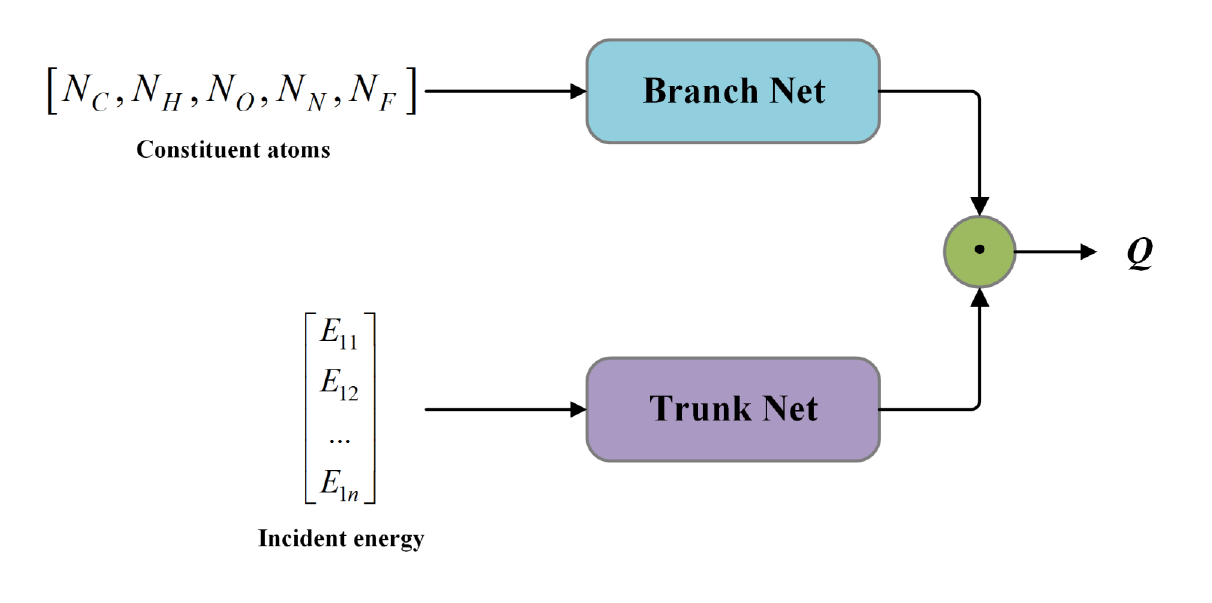}
	\caption{The framework of DeepCSNet for electron-impact ionization cross section}
	\label{fig:fig15}
\end{figure}

\begin{figure}
	\centering
	\includegraphics[width=6.5cm]{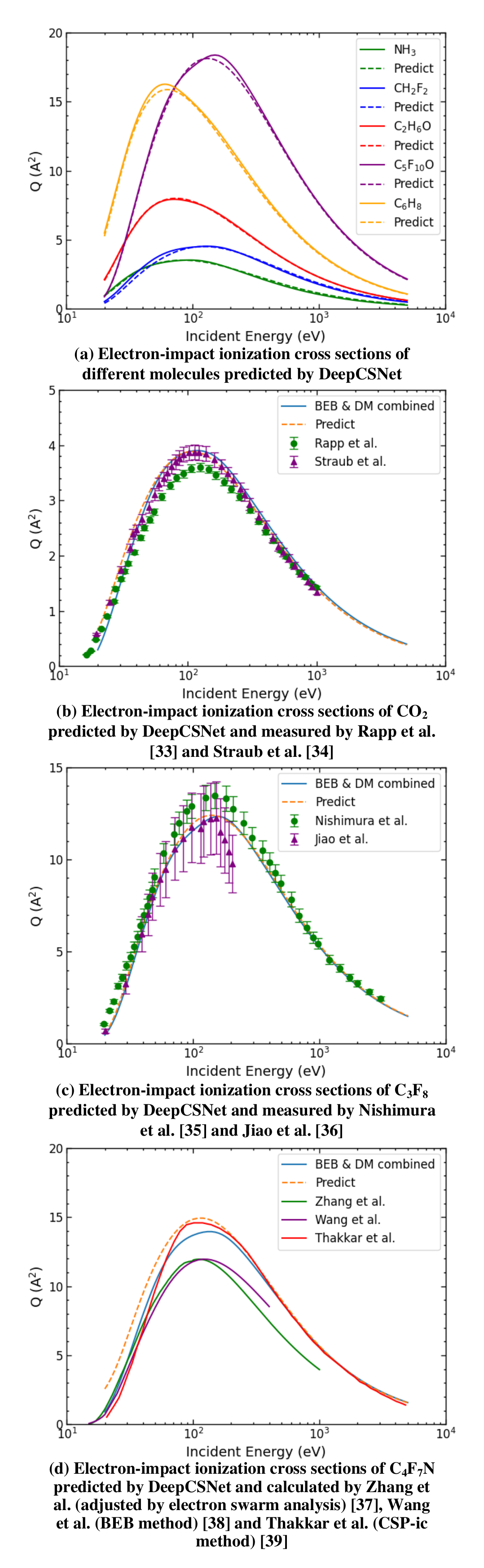}
	\caption{Predictions of electron-impact ionization cross section for different molecules}
	\label{fig:fig16}
\end{figure}

\paragraph{}
An additional test on electron-impact ionization cross sections is conducted. Since no ejected energy is involved in this test, DeepCSNet takes the molecule constituent atoms as the input for the branch net and the incident energy as the input for the trunk net, as shown in Figure \ref{fig:fig15}. Ionization cross sections for 88 different atoms and molecules composed of C, H, O, N, F are considered, with the incident energy ranging from 20 to 4900 eV for each molecule, covering a total of 185 energy points. The exact information about the atoms and molecules can be found in \cite{cite14}. From 20 to 50 eV, the incident energy is sampled every 1 eV, from 50 to 100eV it is 2 eV, from 100 to 1000 every 10 eV, and when it is over 1000 eV, the interval is 100 eV. The data are directly from or generated based on Zhong’s work combining BEB and DM \cite{cite32}. Both the cross section and the incident energy are converted to logarithmic values and normalized to [0.05, 0.95]. Data from 70 molecules are used for training, and the other 18 molecules are used for testing. Both the branch net and the trunk net have 3 hidden layers and 80 neurons per layer. The training setup is the same as before, and the results are shown in Figure \ref{fig:fig16}.

\paragraph{}
The results in Figure \ref{fig:fig16}(a) demonstrate high consistency with the data for unknown molecules, achieving a mean relative $L^2$ error on the test set of 0.074 and the smallest relative $L^2$ error of 0.0070. The curve shape, the value and the position of the peak can be accurately predicted. Figure \ref{fig:fig16}(b) and Figure \ref{fig:fig16}(c) show the comparison between electron-impact ionization cross sections of CO$_2$ and C$_3$F$_8$ predicted by DeepCSNet and collected from experimental measurements. The measurement values adopted for CO$_2$ is recommended by Raju \cite{cite40}. The predictions by DeepCSNet exhibit consistency with the measurements. Figure \ref{fig:fig16}(d) gives the comparison between the results of C$_4$F$_7$N predicted by DeepCSNet and calculated by other methods including electron swarm analysis adjustment \cite{cite37}, BEB method \cite{cite38} and the complex scattering potential-ionization contribution (CSP-ic) method \cite{cite39}. The predicted results display acceptable consistency with the calculations from other works, especially Thakkar’s work \cite{cite39}. These results underscore DeepCSNet's robust generalization ability across different molecules, including large molecules with more than 10 constituent atoms like C$_5$F$_{10}$O. Such generalization capability suggests a promising new approach for cross section calculation. Meanwhile, DeepCSNet exhibits high calculation efficiency. Take C$_5$F$_{10}$O for an example, with an NVIDIA GeForce RTX 3090 GPU, the training process takes about 16min and the inference process takes 2.0s, while the method combining BEB and DM \cite{cite32} takes about 166h with 16 cores Intel Xeon E5-2630 v4 @ 2.20GHz CPU.

\section{CONCLUSION}
\label{sec:sec5}
\paragraph{}
In this paper, DeepCSNet, a deep learning-based method, is proposed for predicting doubly differential cross sections in electron-impact ionization processes. Two configurations of DeepCSNet are introduced and investigated: SMC and MMC. The former predicts DDCS for a specific molecule, while the latter predicts DDCS for different molecules, with the constituent atoms as additional input. Both configurations can successfully predict DDCS and demonstrate generalization ability to unknown incident and ejected energies, as well as unknown incident angles not included in the training set, within or outside the training ranges. In MMC, a comparison between models with a single branch net and with separate branch nets is conducted. DeepCSNet with two branch nets exhibits superior performance in both accuracy and generalization, indicating its potential to predict cross sections at any incident angle or energy range.

\paragraph{}
To further illustrate the generalization abitlity of DeepCSNet for unknown molecules, an additional test on electron-impact ionization cross sections is performed. The results confirm DeepCSNet’s capability to accurately predict cross sections for unknown molecules, including those with more than 10 constituent atoms, highlighting its potential for practical applications.

\paragraph{}
In future work, more information about molecules can be provided to the network, to enhance the prediction accuracy and reasoning. Furthermore, more complex and informative neural networks, such as convolutional neural networks or graph neural networks, can be introduced into the framework to further improve its performance.

\section*{ACKNOWLEDGEMENTS}
\label{sec:acknowledgments}
\paragraph{}
This work was supported in part by the Natural Science Foundation of Jiangsu Province (BK20231427), the National Natural Science Foundation of China (92066106), the Zhishan Young Scholar Project of Southeast University (2242022R40022), and the Fundamental Research Funds for the Central Universities (2242022R40022).

\section*{AUTHOR DECLARATIONS}
\label{sec:author}
\subsection*{Conflict of Interest}
\paragraph{}
The authors have no conflicts to disclose.

\subsection*{Author Contributions}
\paragraph{}
Yifan Wang: Data curation; Software; Investigation; Validation; Visualization; Writing – original draft; Linlin Zhong: Conceptualization; Methodology; Funding acquisition; Supervision; Writing – review \& editing.

\section*{DATA AVAILABILITY}
\label{sec:data}
\paragraph{}
The data that support the findings of this study are available from the corresponding author upon reasonable request.

\bibliographystyle{unsrt}


\end{document}